\documentstyle[prl,aps,floats,twocolumn,psfig]{revtex}

\def\etal{{\it et. al.}}

\begin{document}
\wideabs{
\title{Resistive transition and upper critical field in underdoped YBa$_2$Cu$_3$O$_{6+x}$ single crystals.}
\author{V. F. Gantmakher\thanks{E-mail: gantm@issp.ac.ru} and G. E.
Tsydynzhapov}

\address{Institute of Solid Stale Physics. Russian Academy of Sciences, 142432
Chernogolovka. Moscow Region, Russia}

\author{L. P. Kozeeva and A. N. Lavrov}
\address{Institute of Inorganic Chemistry. Siberian Department of Russian
Academy of Sciences, 630090 Novosibirsk. Russia}
\maketitle

\begin{abstract}

A superconducting transition in the temperature dependence of the $ab$-plane
resistivity of underdoped YBa$_2$Cu$_3$O$_{6+x}$ crystals in the range
$T_c\lesssim30$~K  has
been investigated.  Unlike the case of samples with the optimal level of
doping, the transition width increased insignificantly with magnetic field,
and in the range $T_c\lesssim$13~K it decreased with increasing magnetic field. The
transition point $T_c(B)$ was determined by analyzing the fluctuation
conductivity. The curves of $B_{c2}(T)$ measured in the region
$T/T_c\gtrsim0.1$ did not show a tendency to saturation and had a positive
second derivative everywhere, including the immediate neighborhood of $T_c$.
The only difference among the curves of $B_{c2}(T)$ for different crystal states
is the scales of $T$ and $B$, so they can be described in terms of a universal
function, which fairly closely follows Alexandrov's model of boson
superconductivity.
\end{abstract}
}

\section{Introduction}
\label{intro}

The nature of high-temperature superconductivity is presently one of the most
interesting subjects of the solid state physics. An important topic of
research in this field is the temperature dependence of the upper critical
field $B_{c2}$. In conventional (low-temperature) superconductors, in accordance
with the BCS theory, the curve $B_{c2}(T)$ is described by a universal function
$b_{BCS}(t)$ in terms of reduced variables: the temperature is scaled by the
zero-field transition temperature, $t=T/T_c$, and the magnetic field is
scaled by the product of $T_c$ and the derivative of $B_{c2}(T)$ at $T_c$:
$b =B/[T_c(-dB_{c2}/dT)_{T=T_c}]$.\cite{1} The function $b_{BCS}(t)$ is linear in the neighborhood of $T_c$
and saturates to $b\approx0.7$ at $t=0$. In high-temperature superconductors (HTSC)
the behavior of $B_{c2}(T)$ is radically different. In
Tl$_2$Ba$_2$CuO$_6$\cite{2} and Bi$_2$Sr$_2$CuO$_y$\cite{3} films, and in
K$_{0.4}$Ba$_{0.6}$BiO$_3$ single crystals,\cite{4,5} a positive second
derivative and a sharp increase in $B_{c2}(T)$ at low temperature have been
detected. Similar properties of function $B_{c2}(T)$ have been observed in
other HTSC systems, namely, in YBa$_2$(Cu$_{1-y}$Zn$_y$)$_3$O$_{6+x}$ with a critical temperature
lowered   by   the strong   scattering\cite{6}   and Sm$_{1.85}$Ce$_{0.15}$CuO$_{4-y}$ with
$n$-type conductivity.\cite{7}

HTSC is not the only class of materials where the upper critical field does
not follow the BCS universal function $b_{BCS}(t)$. But, as concerns HTSC, such
deviations are probably present in all materials of the family, and
magnitudes of these deviations are enormous.\cite{2,3} Therefore, there is every
reason to seek fundamental causes of these deviations, which are general for
all HTSC.

Several models have been suggested. Ovchinnikov and Kresin\cite{8} focused attention
on magnetic impurities, which, as they assumed, cause pair breaking and
effectively suppress superconductivity near $T_c$. The tendency to magnetic
ordering at lower temperatures results in a lower spin-flip scattering
amplitude, thus enhancing superconductivity. The presence of magnetic
impurities is a common feature of HTSC, since current carriers in most of
them are due to doping, which generates magnetic defects at the same time.

Spivak and Zhou\cite{9} studied the role of Landau quantization combined with
a random potential. The quantization leads to a higher density of states on
Landau levels, whereas the random potential brings to the Fermi level Landau
sub-levels with opposite spins at points close to one another in space. In
this case, the random potential must satisfy two opposite conditions: its
variation over the coherence length $\xi$ should be larger than the Zeeman
splitting, on the other hand, scattering by this potential should not wipe
away peaks in the density of states. The HTSC structure favors both these
conditions: fluctuations in the concentration of dopants, which are at the
same time scattering centers, should occur even in high quality crystals, but
these scatterers and current carriers are separated in space.

It is possible that there are more fundamental causes of the peculiar shape
of $B_{c2}(T)$ curves that can be put down to an exotic nature of superconductivity
in HTSC. One example is the "bipolaron" or, in a more general approach, the
"boson" model of superconductivity suggested by Alexandrov and Mott.\cite{10} The
model assumes that pairs (charged bosons, e.g., bipolarons) are preformed,
and the superconducting transition consists in Bose-condensation of these
pairs. In the presence of a random potential, the curve of $B_{c2}(T)$ has a
positive curvature. The conventional superconductivity in a Fermi liquid can
transform to the boson superconductivity if the electron--phonon coupling is
strong and the carrier density is low. Again, HTSC materials are good
candidates for realization of such a scenario. Their carrier concentration is
lower than in conventional metals and drops further with decreasing doping
level, whereas the coupling constant $\lambda\gtrsim1$.

Abrikosov suggested for HTSC a model whose central component is a saddle-like
singularity in the electron spectrum. This model predicts, in particular, a
positive curvature of the $B_{c2}(T)$ curve\cite{11} because the problem becomes
effectively one-dimensional due to the saddle point; as a result, the
magnetic field's capability of destroying superconductivity is limited
considerably. In the absence of the paramagnetic limit, the model yields the
divergent function $B_{c2}(T)$, but if the paramagnetic limit is taken into
account, the critical field is limited to a finite value.

The experimental data accumulated over recent years are insufficient for
making an ultimate choice of one of these model. Further research is needed,
and the present paper is a step in this direction. We present an
investigation of the effect of a magnetic field on the resistivity of
YBa$_2$Cu$_3$O$_{6+x}$ single crystals at doping levels below the optimal one. The
aim of this work was to measure the temperature dependence of $B_{c2}$ in this
material at $x$ such that $T_c<30$~K and derive from these data changes in
parameters that control $B_{c2}$ when $T\to0$.

The paper is organized as follows. Section~\ref{review} presents basic
theoretical concepts concerning the superconductor phase diagram in a
magnetic field and the behavior of conductivity around the superconducting
transition point; they are essential in the analysis of experimental data.
Section~\ref{exper} describes sample fabrication techniques and experimental
procedures. Sec.~\ref{results} reports on experimental results. The curves
$\rho(T)$ and their evolution induced by the magnetic field are discussed in
Sec.~\ref{rho}. The derivation of $B_{c2}(T)$ from resistance-versus-temperature
data for HTSC has remained a controversial issue,\cite{12,13} therefore this
topic is given special treatment in Sec.~\ref{bc2}.  Since the transition
broadening induced by magnetic field is insignificant, qualitative
conclusions concerning the behavior of $B_{c2}(T)$ are not affected by the specific
routine employed in determination of the superconducting transition point.
Nonetheless, in determining $B_{c2}(T)$ quantitatively, we analyzed the
fluctuation conductivity in the normal state as a function of temperature.
Section~\ref{discuss} discusses $B_{c2}(T)$ derived from experimental data:  the
curvature of $B_{c2}(T)$ curves proved to be positive throughout the available
temperature range, including the close neighborhood of $T_c$; no signs of
saturation in the low-temperature range have been detected; the experimental
data are compared with existing models.

\section{Basic theoretical concepts.}
\label{review}
\subsection{Phase diagram}

The phase diagram of a type-II superconductor in the $B$--$T$ plane in the
mean--field approximation contains a Meissner region, where magnetic field is
fully ejected from a sample, a mized state region, where a lettice of
Abrikosov's flux lines exists and a normal metal region. These regions are
separated by lines of second order phase transitions:  $B_{c1}(T)$ between
the Meissner and mixed phases and $B_{c2}(T)$ between the mixed state and
normal metal.

Beyond the mean--field approximation, thermal fluctuations of the order
parameter slightly change the phase diagram configuration. Now, it contains a
region of "vortex liquid," where fluctuations change largely the order
parameter phase (which can be interpreted in terms of free motion of
Abrikosov's flux lines), and a region of critical fluctuations close to
$B_{c2}(T)$, where the order parameter amplitude fluctuates and its mean
value changes rapidly with the temperature or magnetic field intensity. There
are superconducting fluctuations above $B_{c2}(T)$ also, but their amplitude
is small and decreases away from the line of $B_{c2}(T)$. The phase
transition to the superconducting state with a long-range order established
occurs on the boundary between the vortex liquid and vortex lattice [melting
line $B_m(T)$], whereas the curve of $B_{c2}(T)$ determined in the mean--field
approximation defines the line of a crossover from the normal metal, where
the order parameter fluctuation amplitude is low, to the vortex liquid, where
the magnitude of the order parameter is almost unity.\cite{14,15,16}

In conventional superconductors, the regions of critical fluctuations and
vortex liquid are quite narrow and essentially unobservable. The melting line
$B_m(T)$ coincides with $B_{c2}(T)$, therefore, the mean--field approximation
adequately describes the phase diagram. In HTSC the situation is different.
Owning to the high critical temperature, small coherence length, and high
anisotropy, fluctuations play a more important part, and the vortex liquid
phase occupies a considerable region of the phase diagram, so $B_m$ and $B_{c2}$ are
separated. Since fluctuations broaden features of field dependencies of
transport and thermodynamic properties at point $B_{c2}$, it is most difficult to
determine this point in experiment. Nonetheless, the value of $B_{c2}(T)$ is
still very important since this is the parameter that controls the behavior
of thermodynamic quantities in the region far from the line of transition,
where the mean--field approximation is valid.

In materials with strong pinning, the phase diagram is further modified: the
pinning destroys the order in the vortex lattice and transforms it to a
vortex glass. The melting line is replaced by the "irreversibility line"
$B^*(T)$, above which vortices are depinned by thermal fluctuations and move
freely even at very low current densities, which results in a finite
resistivity and reversible $dc$ magnetization. Below $B^*(T)$ vortices are
pinned in the low current limit, and the magnetization curve shows a
hysteresis.

\subsection{Resistive transition}
In high-temperature superconductors with optimal doping, curves of $\rho(T)|_B$
form a fan with a common transition onset point, so the positions of the
transition onset are almost independent of the magnetic field.\cite{14,17}
The drop in the resistivity around the transition onset is controlled by the
contribution of superconductive fluctuations to the conductivity. The
characteristic field of fluctuation suppression is $B_{c2}$, hence the shift
of the transition onset should follow the function $B_{c2}(T)$. On the
low-temperature side, the resistivity should vanish
when the vortex motion is frozen. Qualitatively, the line on the $B$--$T$ phase
diagram where the vortex mobility becomes significant is the "irreversibility
line" $B^*(T)$. Thus, the resistive transition is confined by the lines
$B_{c2}(T)$ and $B^*(T)$ and is associated with the vortex liquid region on
the phase diagram so that the fan-like appearance of resistivity curves is
due to broadening of this region with the magnetic field while the line
of $B_{c2}(T)$ is almost vertical.

The breadth of the vortex liquid region, hence the transition width, is
determined by the relation between pinning and fluctuations. The vortex
depinning is favored by the small coherence length $\xi$, high temperatures,
and weak coupling between neighboring superconducting layers of CuO$_2$,
i.e., by the high anisotropy. Variations in the doping level (carrier density
$n$) to both sides from the optimal doping $n_{opt}$, lead to lower $T_c$ and
larger $\xi$.  On the other hand, the anisotropy is stronger at lower doping
and weaker at higher doping levels. The resistivity curves of overdoped HTSC
samples with high carrier densities and low anisotropy are similar to those
of conventional superconductors with  strong pinning.\cite{2,18}

The difference between over- and underdoped states was demonstrated by
comparing La$_{2-x}$Sr$_x$Cu0$_4$ samples with different $x$.\cite{18}
Whereas a magnetic field of $B = 8$~T broadened by 15--20~K the resistive
transition in an underdoped sample with $x=0.08$ and $T_c\approx30$~K, the
transition curve in an overdoped sample with $x=0.20$ and approximately the
same $T_c$ was shifted by magnetic field without changing its shape.\cite{18} This
observation was confirmed by other researchers,\cite{19,20} who also reported
that decreasing the oxygen content in YBa$_2$Cu$_3$O$_{6+x}$ thin films and
single crystals considerably enhances effects originated from vortex motion,
in particular, increases transition broadening in the magnetic field. All
these experiments, however, used samples with $T_c\gtrsim40$~K, and it remained unclear
whether this tendency should persist in the range of low $T_c$.

There is an alternative interpretation of the resistive transition in
cuprates, which attributes most of the change in the resistivity to a phase
transition between the vortex liquid and vortex lattice (vortex glass) at
$B_m(T)$.\cite{14,21,22} In this case, the resistive transition is decomposed into a
resistivity jump on the $B_m(T)$ line [well below $B_{c2}(T)$] and a
crossover on line $B_{c2}(T)$,\cite{21} which can produce only slight changes
in resistivity.

The high conductivity in the normal state of overdoped cuprates might in fact
mask the transition from the normal to vortex liquid state.\cite{2} But changes in
transport characteristics around $B_{c2}$ are evident even in high quality
YBa$_2$Cu$_3$O$_{6+x}$ crystals with optimal doping and very weak pinning.\cite{23}
They should be the much more notable in underdoped samples, whose
conductivity in the normal state is essentially lower.

\section{Experimental}
\label{exper}

YBa$_2$Cu$_3$O$_{6+x}$ single crystals were grown by slow cooling the melt
containing 10.0 to 11.4 wt.\% of YBa$_2$Cu$_3$O$_{6+x}$ and eutectic mixture
of 0.28 BaO and 0.72 CuO as a solvent with subsequent decanting of residual
flux. For our experiments, we selected single crystals without visible signs
of block structure and shaped as plates
20~to 40~$\mu$m thick with areas of several square millimeters. After
oxygenating at 500$^o$C, they had $T_c\approx$90--92~K and fairly narrow
resistive superconducting transitions with $\Delta T<1$~K.

In YBa$_2$Cu$_3$O$_{6+x}$, current carriers (holes) are generated in CuO$_2$
planes as a result of capturing electrons in layers of CuO$_x$ chains. The hole
density depends on the oxygen content $x$ and configuration of oxygen atoms in
chains in CuO$_x$ layers. Consequently, the carrier density in YBa$_2$Cu$_3$O$_{6+x}$
(along with the superconducting transition temperature) can be varied by two
methods: changing the oxygen content and varying its ordering in CuO$_x$ layers.

The technique for changing the oxygen content is the high-temperature
annealing, and it allows one to produce the whole range of states from
antiferromagnetic insulator to optimally doped superconductor. The annealing
temperature at a given partial pressure of oxygen controls the oxygen content
in a crystal and is a convenient technological parameter in processing
superconducting samples.\cite{24} In order to reduce the oxygen content to
$x=0.37$--0.47, we annealed crystals in air at 700--800$^o$C and then quenched them
in liquid nitrogen to prevent exchange of oxygen with the atmosphere during
cooling.

The second technique allows us to vary the carrier density over a relatively
narrow interval by changing the average length of oxygen chains at constant
$x$.\cite{25,26} In chains of finite lengths, there are $q$ oxygen atoms per $q+1$
copper atoms, hence, one has $(q+1)/q$ electrons per oxygen atom. For this
reason, oxygen atoms in shorter chains are less efficient in capturing
electrons from CuO$_2$ planes. The average chain length can change owing to the
high diffusion mobility of oxygen in CuO$_x$ layers at the room temperature and
above. Longer chains have lower energy, but they contribute less to the
entropy, which makes them less preferable at high temperatures. The balance
between these two factors determines the average chain length in equilibrium
(hence, the number of holes) as a function of temperature. The relaxation
time strongly depends on temperature, so rapid cooling freezes the oxygen
configuration, thus fixing the carrier density. In real experiments, we
heated crystals to 120--140$^o$C and then quenched them in liquid nitrogen. This
procedure notably reduced the number of holes in the sample, hence lowered
$T_c$. After that samples could be stored in liquid nitrogen for indefinitely
long times without any changes whatsoever. If a sample was exposed to the
room temperature, the carrier concentration increased gradually owing to
oxygen coagulation in longer chains. This aging process could be monitored
continuously by measuring the sample resistance at a constant temperature and
interrupted at any moment by cooling the sample, thus we could obtain any
intermediate value of $T_c$. The aging of a sample at the room temperature for
several days returns it to its initial equilibrium state. Since all
restructuring processes in the oxygen subsystem proceed at relatively low
temperatures, this method allows one to obtain a sequence of sample states
with minimal differences in configurations of defects and pinning centers.

All in all, we have studied three crystals at several carrier densities in
each. The sample parameters are listed in Table~\ref{table}. The different
crystals are numbered 1 to~3, their states with different oxygen contents are
labeled a and b, and the quenching states are referred to as {\it quenched},
{\it intermediate}, and {\it aged}. The ratio between resistances at the room
temperature and 50~K, when the free path is largely controlled by defect
scattering, is a characteristic of crystal purity. This parameter of sample 2
is a factor of about three higher than in samples 1 and 3. Parameter $B_{sc}$ will
be discussed in Sec.~\ref{discuss}.

\begin{table}
\caption{Samples}
\label{table}
\begin{tabular}{cccccc}
Sample& $\rho_{room}/\rho_{50\,K}$ & $x$ & Quenching degree& $T_c$,~K &
$B_{sc}$,~T\\
\hline
&&&quenched&16.5&3.0\\
1a &3 &0.43 &intermediate &20.5 &3.8 \\
&&&aged &25.5 & 8.9\\
\hline
2a &8 & 0.41& aged& 19& 2.8\\
\hline
2b&10&0.47&quenched&38.5&120\\
&&&aged&44.5&240\\
\hline
3a&3&$\approx$0.37&quenched&0&---\\
&&&aged&6.3&0.61\\
\hline
3b&3&$\approx$0.37&aged&$\approx$3&---\\
\end{tabular}
\end{table}

We measured the resistance in the $ab$ plane using a four-terminal circuit.
Since YBa$_2$Cu$_3$O$_{6+x}$ crystals with low oxygen contents are highly anisotropic,
it is very important that the current be uniformly distributed over the
sample thickness, so that only one component of the resistivity tensor is
measured. To this end, the current contacts were fabricated over the entire
surfaces of two opposite crystal faces. The contacts were made by a silver
paste and fixed by annealing before all thermal manipulations designed to
vary the hole concentration. The resistance was measured by the standard
technique using a nanovolt-range lock-in amplifier at 23~Hz. The probe
current was weak enough to ensure the linear regime and avoid overheating
even at the lowest temperatures. The uncertainty in the geometrical factor
restricted the accuracy of absolute measurements of conductivity to 10--20\%,
nonetheless, note that the geometrical factor of each sample was the same in
all conducting states.

Most of experiments were performed in a cryostat with a ${}^3$He pumping
system, which allowed us to vary the temperature between 0.3--300~K.\cite{27}
At temperatures of 0.3--1.2~K the sample was immersed in liquid ${}^3$He, at
higher temperatures it was in the ${}^3He$ atmosphere at a pressure of several torr
serving as a heat-exchange gas. The temperature was measured by a carbon
resistance thermometer calibrated by a reference platinum thermometer, $^4$He
vapor pressure, and cerium-magnesium nitrate in appropriate temperature
ranges. The magnetic field of up to 8.25~T was applied along the $c$-axis.

Sample 3b in the aged state with low $T_c$ was tested in a dilution
refrigerator at temperatures down to 30~mK and magnetic fields of up to 14~T.

\section{Results}
\label{results}
\subsection{Temperature dependence of resistivity}
\label{rho}
In our experiments on samples with $T\ge30$--35~K (sample 2b quenched anh
aged), we record fans of $\rho(T)|_B$ curves, similar to those reported by
other authors.\cite{19,20} In
samples with lower $T_c$, the effect of magnetic field on the resistive
transition is radically different, and in this publication we concentrate on
these effects, namely, the behavior of YBa$_2$Cu$_3$O$_{6+x}$ samples in
states with $T_c\le30$~K (samples la, 2a, 3a, and 3b in all quenching states). In
these samples, magnetic field shifts the transition without a notable
broadening (Fig.~\ref{f1}), which indicates that the effect of vortex motion on the
shape of transition curve no longer dominates. Nonetheless, the shape of the
transition curve is affected by the magnetic field, and one can see on curves
of temperature derivatives $\partial\rho/\partial T$ plotted on the right of
Fig.~\ref{f1} that these changes are nonmonotonic. Since the normal state
resistivity is almost constant with temperature, the peak amplitude on the
derivative curve is inversely proportional to the resistive transition width.
These graphs clearly show that, irrespective of $T_c$ ($\le$30~K), the transition
width is maximum at about 13--14~K. If the zero-field $T_c$ is higher, the
transition first shifts to lower temperatures with magnetic field and
broadens (Fig.~\ref{f1}a). Then, below 13--14~K, the transition narrows concurrently
with its shift to lower temperatures. If $T_c$ is initially lower than 13--14~K
(Fig.~\ref{f1}b and~\ref{f1}c), the transition is narrowed by magnetic field concurrently
with its shift to lower temperatures from the start, and the slope of the
transition curve in magnetic field becomes steeper than at zero field.

The comparison between samples la and 2a demonstrates that the nonmonotonic
change in the transition width with magnetic field is a reproducible property
and is little affected by the sample quality. The superconducting transition
temperatures of these two samples were driven to one value by annealing
(Fig.~\ref{f1}b and~\ref{f1}c), but their parameters in the normal state were
notably different.  Sample 2a contained less impurities and structural
defects, as a result, its resistivity around $T_c$ was twice as small (Fig.~\ref{f1}),
it dropped more rapidly in the process of cooling from the room temperature
to 50~K (Table~\ref{table}) and showed a smaller increase in the range of lower
temperatures. Nonetheless, irrespective of all these differences, both the
transition shift rate in magnetic field and the evolution of transition
curves of these samples are similar. Narrowing of the resistive transition in
an underdoped YBa$_2$Cu$_3$O$_{6+x}$ with increasing magnetic field in this
temperature range was detected by Seidler \etal.\cite{25} but, since their
measurements were presented in a different form, it is difficult to compare
them directly with our results.

Such a behavior of transition curves is observed for all samples with $T_c\ge6$~K.
In states with lower transition temperatures, we were not able to achieve
sufficiently narrow transitions at zero magnetic field to measure $T_c$ and
transition width. Therefore, the measurement data for sample 3b will be given
and discussed separately in Sec.~\ref{discuss}.

\subsection{Derivation of $B_{c2}(T)$ from resistance measurements}
\label{bc2}
The absence of the notable transition broadening in magnetic field in
YBa$_2$Cu$_3$O$_{6+x}$ samples with low $T_c$ indicates that, unlike samples with
$T_c\ge30$--35~K, they have a narrower region of the "vortex liquid" on the phase
diagram. The transition width, however, is not so small that it could be
neglected in determining $B_{c2}(T)$. Since the point $B_{c2}$ is not
marked by a sharp feature on curves of $\rho(T)$, there is a good reason to
determine this point by fitting a theoretical curve describing the crossover
between the normal metal and vortex liquid to the experimental data. In
developing this approach, let us consider the sample conductivity as a sum of
the normal and fluctuation components: $\sigma(T)=\sigma_n(T)+\sigma_{fl}(T)$.

\begin{figure}
\vbox{
\psfig{file=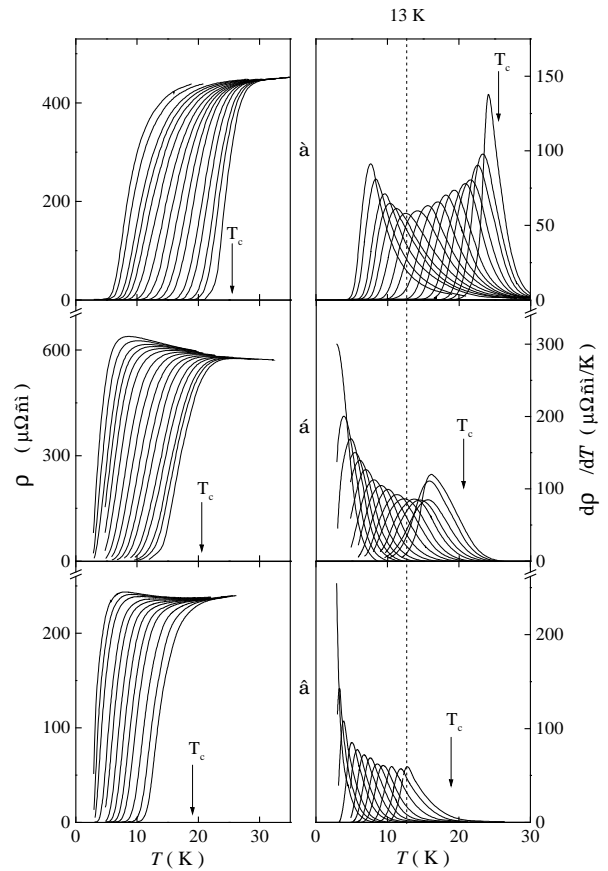,width=\columnwidth,clip=}
\caption{Conductivity $\rho_{ab}$ (on the left) and its derivative (on the
right) at various magnetic fields aligned with the $c$-axis.\protect\\ (a) Sample la in
the aged state; applied fields (from right to left): 0, 0.06, 0.12, 0.23,
0.35, 0.6, 0.8, 1.2, 1.6, 2.2, 3.0, 3.8, 4.6, 5.5, 6.7, and 8.2~T.\protect\\ (b) Sample
la in the intermediate state; applied fields: 0, 0.06, 0.12, 0.23, 0.35, 0.6,
0.8, 1.2, 1.6. 2.2, 3.0, 3.8. 4.6, 5.5, 6.7. and 8.2~T.\protect\\ (c) Sample 2a in the
aged slate; applied fields:  0.12, 0.23, 0.5, 0.8, 1.2, 1.6, 2.2, 3.0, 3.8,
5.5, 6.7, and 8.2~T.}
\label{f1}
}
\end{figure}

The fluctuation conductivity $\sigma_{fl}$  in quasi-two-dimensional systems
in zero field is usually described by the Lawrence--Doniach formula:
\begin{equation}
\label{LD} \sigma_{fl}=\frac 1{16}\frac{e^2}{\hbar d
\epsilon}[1+(2\xi_c(0)/d)^2\epsilon^{-1}]^{-1/2}; \;\;\;\;
\epsilon\equiv\ln{T/T_c},
\end{equation}

where $d$ is the interplane separation. Friedman \etal\cite{28} show that, even in
analyzing optimally doped YBa$_2$Cu$_3$O$_{6+x}$, crystals with the resistivity
anisotropy no higher than 30--100, one can neglect the factor in brackets
which takes into account effects of the third dimension and use
Aslamazov--Larkin's expression for two dimensions:

\begin{equation}
\label{AL}
\sigma_{fl}=\frac 1{16}\frac{e^2}{\hbar d}\epsilon^{-1}.
\end{equation}

In oxygen deficient crystals, the anisotropy is up to
(5--10)$\cdot10^3$,\cite{24} therefore Eq.(\ref{AL}) is {\it a fortiori}
valid throughout the temperature range in question, except the neighborhood
of $T_c$.

There is no consistent theoretical description of $\sigma_{fl}(T,B)$ in
nonzero magnetic field for arbitrary $B_{c2}(T)$. Ullah and Dorsey\cite{16} analyzed
$\sigma_{fl}$ in a system with strong fluctuations in magnetic field and suggested a
scaling expression for the fluctuation conductivity, which is often used in
describing the resistive transition and determining $B_{c2}(T)$ of cuprate
superconductors.\cite{29,30,31} Since their approach is based on the mean--field
approximation and assumes a linear dependence $B_{c2}(T)$ near $T_c$, it does
not apply when $B_{c2}(T)$ is strongly nonlinear. (It will be shown below
that this is the case in our samples.) Nonetheless, in the region well above
$T_c(B)$ ($\epsilon_B\agt0.1$), where Gaussian fluctuations dominate, a formula
similar to that suggested by Aslamazov and Larkin can be used:
\begin{equation}
\sigma_{fl}\propto \epsilon_B^{-1}
\mbox{,\qquad  $\epsilon_B=\ln\frac{T}{T_c(B)}$}
\label{AL-B}
\end{equation}
in both zero and finite magnetic fields (see Ref.\onlinecite{16} and references therein).
Here $T_c(B)$ is the functional inverse of $B_{c2}(T)$. This formula also assumes,
generally speaking, a linear dependence $B_{c2}(T)$, but a possible change in
the exponent of this function should lead only to a small systematic shift of
the resulting curve $T_c(B)$.

\begin{figure}
\vbox{
\psfig{file=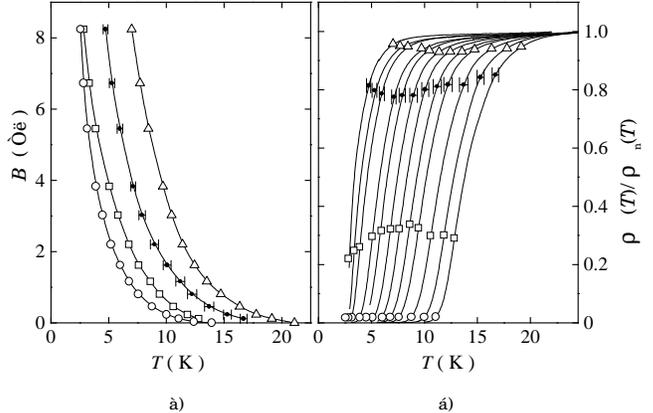,width=\columnwidth,clip=}
\caption{Characteristic points of the superconducting transition in sample 2a plotted
in (a) $B$--$T$ and (b) $\rho$--$T$ planes (the resistivity is normalized to function
$\rho_n(T)$ in Eq.(\protect\ref{sign}), which was used in determination of the fluctuation
conductivity): (open circles) "irreversibility line", $\rho=5$~$\mu\Omega$cm;
(squares) peak of derivative $\partial\rho/\partial T$; (full circles)
$B_{c2}(T)$; (triangles) "transition onset" ${\partial\sigma_{fl}}/{\partial T}=10^2$~$\Omega^{-1}$cm$^{-1}/$K.
}
\label{f2}
}
\end{figure}

In contrast to the case of optimal doping, the normal conductivity in our
samples is low, of order of $e^2/\hbar d$ (Fig.~\ref{f1}), if $d$ is assumed
to be of order of the lattice constant, 11.7~\AA. Simple estimates based on the
Aslamazov--Larkin formula (\ref{AL}) with a reasonable value of $d$ indicate that the
contribution of fluctuations, $\sigma_{fl}$, should be several percent of $\sigma_n$ even at
$\epsilon_B\gtrsim0.5$. This makes determination of $\sigma_n(T)$ more
difficult. The difficulties are exacerbated by the fact that the normal state
resistivity has a minimum in the region of 30--40~K and increases at lower
temperatures. Therefore, we decide to select {\it a priori} the function
$\sigma_n(T)$ with several fitting parameters. The fitting to experimental
data is performed by varying all parameters in both $\sigma_{fl}(T)$ and
$\sigma_{n}(T)$.\cite{29} This procedure could hardly produce sensible
results if each curve $\rho(T)$ were described by a different set of
parameters.  Fortunately,   the magnetoresistance   of YBa$_2$Cu$_3$O$_{6+x}$
crystals in the discussed region of fields and temperatures is negligible in
the normal state, i.e., the shape of $\sigma_{n}(T)$ is constant with the
magnetic field.

Our previous investigations of YBa$_2$Cu$_3$O$_{6+x}$ single crystals near
the boundary of the superconducting region of the phase diagram\cite{32}
revealed that the normal resistivity of such samples at $T<20$~K is well
described by a logarithmic function. In a broader temperature range
(0.5~K$<T< 150$~K) the conductivity is very closely described by the
empirical function
\begin{equation}
\sigma_n(T)=\rho_n^{-1}=[\alpha-\beta \log T+\gamma T]^{-1}
\label{sign}
\end{equation}
This function with three fitting parameters is used in processing our
experimental data.

By approximating the conductivity in zero magnetic field by a sum of $\sigma_{n}(T)$
from Eq.(\ref{sign}) and $\sigma_{fl}(T)$ from Eq.(\ref{AL}), $T_c$ and $d$
being fitting parameters, along with $\alpha$, $\beta$ and $\gamma$, we
obtain reasonable values $d=8$--15~\AA, which are in fair agreement with the
YBa$_2$Cu$_3$O$_{6+x}$ lattice constant along the $c$-axis. This indicates that
the Aslamazov--Larkin formula yields a correct estimate of the fluctuation
conductivity in CuO$_2$ layers and its application is justified. The normal
conductivity is fitted so as to obtain the best approximation of the
fluctuation conductivity throughout the range of magnetic field. Nonetheless,
the uncertainty in the normal resistivity was quite considerable. It turned
out, however, that calculations of the transition temperature are little
affected by admissible variations in $\sigma_{n}(T)$. The resulting
uncertainties in the transition temperature are shown in Fig.~\ref{f2}.

This procedure enable us to derive $B_{c2}(T)$ in the mean--field
approximation from our measurements. Since the resulting curve of $B_{c2}(T)$
is nonlinear and it casts doubt on the applicability of Eq.(\ref{AL-B}), we deem it
necessary to demonstrate that, on the qualitative level, the shape of the
$B_{c2}(T)$ curve is not affected by subtleties of the data processing, owing
to the absence of considerable transition broadening. Figure~\ref{f2}a shows
the curve of $B_{c2}(T)$ for sample 2a, along with its other characteristic
fields, namely, the "irreversibility line" determined at
$\rho=5$~$\mu\Omega$cm, positions of the peak of derivative
${\partial\rho}/{\partial T}$, and the line of "transition onset," which was
defined as a point were ${\partial\sigma_{fl}}/{\partial
T}=10^2$~$\Omega^{-1}$cm$^{-1}/$K. These lines are plotted in the $B$--$T$
diagram in Fig.~\ref{f2}a, and Fig.~\ref{f2}b shows positions of these points on the
transition curves. (It is noteworthy that the values of $B_{c2}$ are fairly close
to those which would be obtained by defining the transition point at a
constant resistivity level $\rho/\rho{n}=0.8$.) Figure~\ref{f2}a clearly shows that all
curves in the $B$--$T$ plane have positive curvature throughout the range of
studied magnetic fields, including the region of low fields. This leads us
to a conclusion that, even if the data processing procedure yields erroneous
values of $B_{c2}$, the temperature dependence of this parameter is
qualitatively correct. Our further analysis, however, will be based on the
values derived from measurement data for the fluctuation conductivity.

\subsection{Universal temperature dependence of the upper critical field}
\label{discuss}

\begin{figure}
\vbox{
\psfig{file=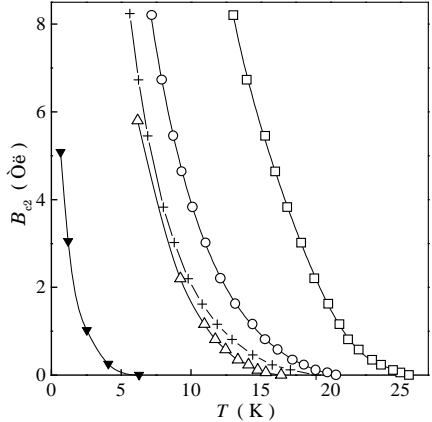,width=\columnwidth,clip=}
\caption{
Temperature dependence $B_{c2}(T)$ in different states. Data for
sample la in (empty triangles) quenched, (open circles) intermediate, and
(squares) aged states; aged states of (crosses) sample 2a and (full inverted
triangles) sample 3a.}
\label{f3}
}
\end{figure}

\begin{figure}
\vbox{
\psfig{file=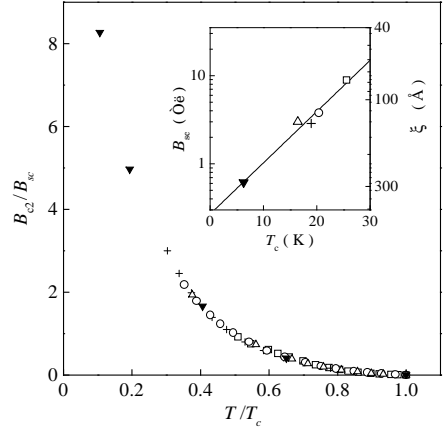,width=\columnwidth,clip=}
\caption{Dependencies $B_{c2}(T)$ for different samples reduced to the
universal function $b_{sc}(t)$ using variables (\protect\ref{Scaling}). The
notation is the same as in Fig.~\protect\ref{f3}. The inset plots the parameters $B_{sc}$
(left-hand axis) and correlation length $\xi_0$ calculated by
Eq.~\ref{Al_red} (right-hand axis).}
\label{f4}
}
\end{figure}

Measurements of $B_{c2}(T)$ in three samples and five different states (all
the states of samples la and 2a and the aged state of sample 3a) are given in
Fig.~\ref{f3}. It turned out that the curves for all the states can be brought to
coincidence by varying the scales of the magnetic field and temperature,
i.e.,

\begin{equation}
B_{c2}=B_{sc}b_{sc}(t), \qquad t=\frac{T}{T_c},
\label{Scaling}
\end{equation}

where $B_{sc}$ is the parameter characterizing the state and $b_{sc}(t)$ is a
universal function (Fig.~\ref{f4}). Function $b_{sc}(t)$ contains an arbitrary numerical
factor. In Fig.~\ref{f4} parameter $B_{sc}$ is defined as $B_{c2}$ at a specific reduced
temperature equal for all samples, namely, $T_c/2$, i.e., the curves of
$B_{c2}(T)$ were brought to coincidence at two points, namely, at $t=1$ and
$t=0.5$. The values of $B_{sc}$ for different states are listed in
Table~\ref{table} and plotted in the inset to Fig.~\ref{f4} as a function of the
zero-field transition temperature. These points lie on one smooth curve, even
though they are derived from measurements of the three different samples. The
characteristic scale of magnetic field decreases (accordingly, the coherence
length increases) with decreasing doping level more rapidly than $T_c$, i.e.,
$B_{sc}$ is a superlinear function of $T_c$. This may be the main cause of the
narrowing of the vortex-liquid region in the $B$--$T$ phase diagram. As a result,
YBa$_2$Cu$_3$O$_{6+x}$ crystals with a high degree of underdoping with $T_c\lesssim30$~K do not
display notable broadening of the resistive transition due to magnetic field.

The function $B_{c2}(T)$ was measured on sample 2b in a
very narrow temperature range, $T/T_c\gtrsim0.9$, owing to the limit on
available magnetic fields. Its second derivative in
this interval is also positive and all measurements of $B_{c2}(T)$ can be
fitted to function (\ref{Scaling}). But, since no data for lower temperatures are
available and the expected critical fields are very high, the measurements of
sample 2b have not been analyzed in this context.

Function $b_{sc}(t)$ is much different from $b_{BCS}(t)$. First, it has no linear
section near $t=1$. This statement relies on Eq.~(\ref{Scaling}), since for each
$B_{c2}(T)$ curve the limited precision allows one to draw a straight line of
a small slope in the region within 1--2~K near $T_c$, but if we consider the
samples with higher $T_c$, this linear region would be more narrow, and the
slope of function $b_{sc}(t)$ at $t= $ is smaller, which leads us to a conclusion
that the universal curve has no linear section near $t=1$.

\begin{figure}
\vbox{
\psfig{file=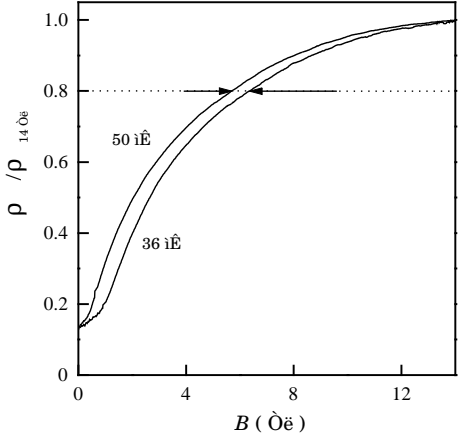,width=\columnwidth,clip=}
\caption{Reduced resistivity of sample 3b in the aged state at temperatures
of 50~and 36~mK as a function of magnetic field. The arrows indicate the
difference between magnetic fields at which the curves cross the level
$\rho/\rho_{14\,T}=0.8$.}
\label{f5}
}
\end{figure}

Second, $b_{sc}$ continues to rise as $t\to0$. Figure~\ref{f4} shows this tendency in
the region down to $t = 0.1$. In order to test the range of lower $t$, we
investigated sample 3b with $T\approx 3$~K at millikelvin temperatures. Its
transition curve is too wide to determine quantitatively $T_c$ and $B_{c2}(T)$.
Nonetheless, the measurements yield important qualitative information.
Figure~\ref{f5} shows the sample resistance versus magnetic field obtained at
temperatures of 50 and 36~mK normalized to the resistance at a magnetic field
of 14~T. It is clear that a drop in temperature shifts the magnetoresistance
curve to lower fields, i.e., $B_{c2}(T)$ still grows with decreasing
temperature even at $T/T_c\sim0.01$. We can obtain the following estimate: on
the level $\rho/\rho_n=0.8$, which approximately corresponds to $B_{c2}(T)$ according
to Fig.~\ref{f2}b, the magnetic field increases by 0.6~T; this yields a derivative
of 40~T/K. Unfortunately, we cannot plot these points in Fig.~\ref{f4} for
the lack of $T_c$ and $B_{sc}$.

This observation of $B_{c2}(T)$ increasing even at very low temperatures is
in accord with measurements of other materials, e.g.,
Tl$_2$Ba$_2$CuO$_6$,\cite{2} where the upper critical field continues to grow
at temperatures down to $T/T_c = 0.001$.

Our data indicate that function $B_{c2}(T)$ in underdoped
YBa$_2$Cu$_3$O$_{6+x}$ is
nonlinear in the neighborhood of $T_c$, and $(\partial B/\partial T)|_{T_c}
=0$. This conclusion contradicts most theoretical models based on the BCS
model or the Ginzburg--Landau functional, which either predict a linear
behavior of this curve near $T_c$ or assume its existence {\it a priori}.
This issue was not discussed in previous publications of experimental
investigations,\cite{2,3,4,5,6,7} but they all reported very low, if not
zero, values of  $(\partial B/\partial T)$  at $T_c$.

The increase in the critical field owing to weakening of the spin-flip
scattering predicted by Ovchinnikov and Kresin\cite{8} should occur in the
range of low temperatures, so it leaves the linearity of $B_{c2}(T)$ near $T_c$.
essentially unaffected. The mechanism suggested by Spivak and Zhou\cite{9}
is effective only in high magnetic fields, where Landau quantization is
significant, i.e., it also should not affect $B_{c2}(T)$ near $T_c$.
Abrikoso\cite{11} derived $B_{c2}(T)$ from the Ginzburg--Landau functional
based on his model, which leads, naturally, to a linear dependence of $B_{c2}$ in
the first order in $1-t$.

The nonlinearity of $B_{c2}(T)$ near $T_c$ follows at present only from the
model of bipolaron superconductivity\cite{10,33} which yields positive
curvature of $B_{c2}(T)$ for a charged Bose-liquid in a localizing potential,
this throughout the entire temperature range. At temperatures that are not
overly low, the model predicts\cite{33}

\begin{eqnarray}
B_{c2}(T)=B_d^*\left(\frac{T_c}{T}\right)^{3/2}
		\left[1-\left(\frac{T}{T_c}\right)^{3/2} \right]^{3/2},    \nonumber \\
\quad B_d^*=\frac{\Phi_0}{2\pi\xi_0^2}\left(1-\frac{n_L}{2n}\right)^{1/2}
\label{Al_red}
\end{eqnarray}

\begin{figure}
\vbox{
\psfig{file=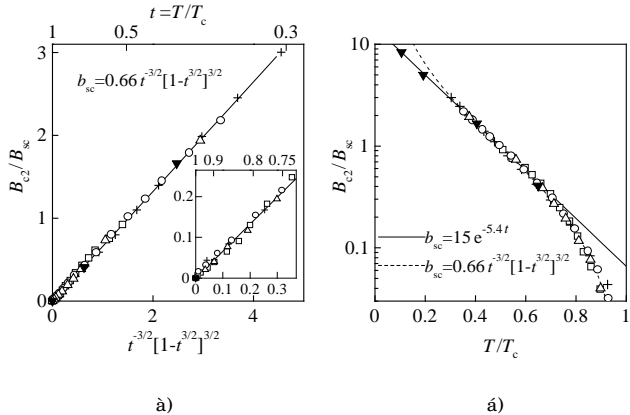,width=\columnwidth,clip=}
\caption{
Function $b_{sc}(t)$ plotted in different coordinates: (a) the coordinates are
selected in accordance with the boson model, Eq.~(\ref{Al_red}); the inset shows the
section close to $t=T/T_c=1$ on the extended scale; (b) semilogarilhmic
coordinates; the dashed line follows Eq.~(\ref{Al_red}).}
\label{f6}
}
\end{figure}

Here $\xi_0$ is the correlation length and $n_L/2n$ characterizes the random
potential. Equation (\ref{Al_red}) defines a universal function in reduced variables
without free parameters. The only normalization parameter $B^*_{d}$ corresponds to
parameter $B_{sc}$ introduced in Eq.~(\ref{Scaling}). It follows from
Eq.~(\ref{Al_red}) that $B^*_{d}=0.68B_{sc}$. Comparison between our data and
calculations by Eq.~(\ref{Al_red}) (Fig.~\ref{f6}a) shows excellent agreement in the region
$T/T_c>0.3$. At lower reduced temperatures experimental points deviate from the
theoretical curve, but note that in this range we have only measurements of
one state (sample 3a aged).

The factor $(1-n_L/2n)^{1/2}$ in Eq.~(\ref{Al_red}) is unknown, but, since neither in state 3a
nor in state 3b have we detected a reentrant behavior of $B_{c2}(T)$
predicted by Alexandrov,\cite{33} it should be rather close to unity. Assuming this,
we can derive from Eq.~(\ref{Al_red}) the correlation length $\xi_0$
(Fig.~\ref{f4}, right-hand axis in the inset). The length $\xi_0$ varies between
70~and 300~\AA. The notable increase in the correlation length may be the main
cause of the narrowing of the vortex liquid region on the $B$--$T$ diagram.

In the low-temperature region $0.1<t<0.6$ the function $b_{sc}(t)$ can be
empirically described by the exponential

\begin{equation}
b_{sc}=b_0\exp(-t/t_0)
\label{Exp(T)}
\end{equation}

with parameters $b_0=15$ and $t_0=5.4$ (Fig.~\ref{f6}b). Such an
unusual temperature dependence in the region of low temperatures was detected
previously\cite{25,34} in measurements of the irreversibility line $B^*(T)$.
We suppose that the region of vortex liquid in phase diagrams of samples with
low $T_c$ is a very narrow strip between $B^*$ and $B_{c2}$, hence $B^*$
closely follows $B_{c2}(T)$, especially at low temperatures.

\section{Conclusions}
Our investigation has supplemented the list of materials displaying anomalous
temperature dependence of the upper critical   field   $B_{c2}(T)$   with
underdoped   cuprate YBa$_2$Cu$_3$O$_{6+x}$. We have studied samples with
different carrier concentrations and $T_c$ ranging between 6 and 30~K.
Throughout the studied temperature range, the curve of $B_{c2}(T)$ for these
samples have positive curvature and does not saturate at low temperatures.
The curves for states with different $T_c$ can be brought to coincidence in
reduced coordinates $T/T_c$ and $B/B_{sc}(T_c)$. A fundamental feature of the
universal function $b_{sc}(t)$ obtained in this manner is the tendency of its
first derivative $\partial B_{c2}/\partial T$ to zero as $T\to T_c$. Such a
behavior can be interpreted at present only in terms of the model\cite{10,33}
treating the superconducting transition as Bose-condensation of preformed
pairs. Other models designed to interpret the anomalous shape of the
$B_{c2}(T)$ curve predict a linear temperature dependence of $B_{c2}$ near
$T_c$.

In the low-temperature range $T/T_c<0.3$, experimental points deviate from
function (\ref{Al_red}). On the other hand, measurements in the temperature
interval between the lowest accessible values and $t\approx0.6$ follow
function (\ref{Exp(T)}). The combination of Eqs.~(\ref{Al_red})
and~(\ref{Exp(T)}) analytically describes function $b_{sc}(t)$.

However, the "universality" of function $b_{sc}(t)$ is limited. We tested
this function on our measurements of K$_{0.4}$Ba$_{0.6}$BiO$_3$,\cite{5} and
the experimental curve after renormalization of variables according to
Eq.~(\ref{Scaling}) was different from function $b_{sc}(t)$ plotted in Fig.~\ref{f4}.

We are indebted to V.T.~Dolgopolov and A.A.~Shashkin for the opportunity to
conduct low-temperature measurements in the dilution refrigerator.

This work was supported by RFBR-PICS (Grant 98-02-22037), by RFBR-INTAS
(Grant 95-02-302), and by the Statistical Physics Program sponsored by the
Russian Ministry of Science and Technology.

\end{document}